\documentclass[graybox]{svmult}

\usepackage{bm}
\usepackage{mathptmx}       
\usepackage{helvet}         
\usepackage{courier}        
\usepackage{type1cm}        
\usepackage{makeidx}         
\usepackage{graphicx}        
\usepackage{multicol}        
\usepackage[bottom]{footmisc}
\usepackage{amsmath}
\usepackage{amssymb}
\usepackage{multirow}
\usepackage{bm}
\usepackage{float}
\usepackage{url}
\usepackage[table]{xcolor}
\usepackage{blkarray}
\usepackage{relsize}

\usepackage{tikz}
\usetikzlibrary{fit}
\tikzset{%
  highlight/.style={rectangle,rounded corners,fill=red!15,draw,fill opacity=0.5,thick,inner sep=0pt}
}
\newcommand{\tikzmark}[2]{\tikz[overlay,remember picture,baseline=(#1.base)] \node (#1) {#2};}
\newcommand{\Highlight}[1][submatrix]{%
    \tikz[overlay,remember picture]{
    \node[highlight,fit=(left.north west) (right.south east)] (#1) {};}
}

\makeindex             

\begin{document}
\title*{Maxima Units Search (MUS) algorithm: methodology and applications}
\subtitle{\emph{ Algoritmo Maxima Units Search (MUS): metodologia e applicazioni}}

\author{Leonardo Egidi, Roberta Pappad\`{a}, Francesco Pauli, Nicola Torelli}

\institute{ Roberta Pappad\`{a}, Francesco Pauli, Nicola Torelli \at  Dipartimento di Scienze Economiche, Aziendali, Matematiche e Statistiche, `Bruno de Finetti', Universit\`{a} degli Studi di Trieste, Via Tigor 22, 34124 Trieste, Italy, 
\email{rpappada@units.it}, \email{francesco.pauli@deams.units.it}, \email{nicola.torelli@deams.units.it} \and Leonardo Egidi \at  Dipartimento di Scienze Statistiche, Universit\`{a} degli Studi di Padova,  Via Cesare Battisti 241, 35121 Padova, Italy, \email{egidi@stat.unipd.it} 
}
\maketitle

\abstract{An algorithm for extracting identity submatrices of small rank and pivotal units from large and sparse matrices is proposed. The procedure has already been satisfactorily applied for 
solving the label switching problem in Bayesian mixture models. Here we introduce it on its own and explore possible applications in different contexts.}
\abstract{\emph{Viene presentato un algoritmo per estrarre sotto-matrici identiche di piccolo rango e unit\`{a} pivotali da una matrice di rango superiore. La procedura \`{e} stata gi\`{a} applicata con buoni risultati per risolvere il cosiddetto problema di `label switching' in modelli mistura Bayesiani: la introduciamo qui come metodo a s\`{e} stante e ne esploriamo ulteriori possibili applicazioni.}}

\section{Introduction}
\label{sec:1}

Identifying and extracting identity matrices of small rank with  given features from a 
larger, possibly sparse, matrix could appear just of theoretical interest. 
However, investigating the structure of a given sparse 
matrix is not only of theoretical appeal but can be useful for a wide variety of practical problems and for statistics. 

This kind of matrix appears in clustering ensembles methods, which combine data partitions of the same dataset in order to
obtain  good data partitions even
when the clusters are not compact and well separated. See, for instance, \cite{fred2002data} where multiple partitions of the same data 
(an ensemble) are generated changing the
number of clusters and using random cluster initializations 
within the K-means algorithm. Another situation where the global number of zeros of a matrix has a relevant role is in analyising the structure of a matrix of factor loadings; \cite{kaiser1974index} introduces and
formulates a statistical index in order to assess how good is the solution based on a factor analysis.

Matrices with a similar structure and for which the sparseness has to be taken into account appear in the so-called cost's optimization theory. \cite{kuhn1955hungarian} 
builds the well-known Hungarian method, which uses the zeros matrix 
elements for finding an optimal assignment
for a given cost matrix; \cite{munkres1957algorithms} presents a generalization of  
such algorithm  and an application to a transportation problem. 

In this paper we discuss the so-called Maxima Units Search algorithm (hereafter MUS). It has been introduced in \cite{pauli2015relabelling} and used in the context of the label switching problem (\cite{jasra2005markov}, \cite{stephens2000dealing}). In Bayesian estimation of finite mixture models label switching arises since 
the likelihood is invariant to permutations of the mixture components. The MUS procedure has proved to be useful in detecting some specific units---one for each mixture component---called pivots, from a large and sparse similarity matrix representing an estimate of the probability that pairs of units belong to the same group. 

The MUS algorithm is then more generally aimed at identifying for a given partition of the data those units that are not connected with a large number of units selected from the other groups.

A formal description of the MUS algorithm is provided and discussed. In fact, we argue that the proposed approach is of a broader interest and can be used for different purposes especially when the considered matrix presents a non-trivial number of zeros.

In Sect.~\ref{sec:meth} we introduce the notation, the algorithm  and the main quantities of interest. A simulation study conducted for exploring the sensitivity of the algorithm to the choice of some parameters is presented in Sect.~\ref{sec:sim_study}. Possible applications are illustrated in Sect.~\ref{sec:appl}: in the first example we report the pivotal identification mentioned above, which represents the initial motivation for the procedure. Finally the method is employed to study a small dataset concerning tennis players' abilities. Sect.~\ref{sec:concl} concludes.

\section{The methodology}
\label{sec:meth}

Let us consider a symmetric square matrix $C$ of dimensions $N \times N$ containing a non-negligible number of zeros and suppose that each row's---or equivalently column's---index represents a statistical unit. Moreover, let us suppose that such $N$ units either naturally belong to $K$ different groups or have been preliminarily clustered into them, for instance via a suitable clustering technique. 

For some practical purposes an example of which will be given in Sect.~\ref{sec:appl}, we may be interested in detecting those units---one for each group---whose corresponding rows have more zeros than the other units. We preliminarily refer to these units as the \textit{maxima} units. More precisely, the underlying idea is to choose as maxima those units $j_{1},...,j_{K}$ such that the $K \times K$ submatrix of $C$, $S_{j_{1},...,j_{K}}$ with only the $j_{1},...,j_{K}$ rows and columns has few, possibly zero, non-zero elements off the diagonal (that is, the submatrix $S_{j_{1},...,j_{K}}$ is identical or nearly identical). Note that an identity submatrix of the given dimension may not exist.
From a computational point of view, the issue is non-trivial and involves a global search row by row; as $N$, $K$ and the number of zeros within $C$ increase, the procedure becomes computationally demanding. 




\begin{figure}
\begin{minipage}[c]{\textwidth}
$S.\ \mathbf{1}$
  \[ 
  \mathsmaller{\mathsmaller{\mathsmaller{
   \bordermatrix{
  	& \textcolor{red}{ 1} & \textcolor{red}{2} &\textcolor{red}{3} & \textcolor{green}{4} & \textcolor{green}{5} & \textcolor{green}{6} & \textcolor{blue}{7} & \textcolor{blue}{8} & \textcolor{blue}{9}  \cr
  	\textcolor{red}{ 1} & 1 & 1 & 1 & 1 & 1 &0 &1&0&0 \cr
    \textcolor{red}{ 2} & 1 & 1 & 0 & 0 & 1 &0 &1&0&0  \cr
    \textcolor{red}{ 3} & 1 & 0 & 1 & 0 & 1&0 &1&0&0  \cr
    \textcolor{green}{ 4}& 1 & 0 & 0 & 1 & 1&0 &1&1&0  \cr
    \textcolor{green}{ 5} & 1 & 1 & 1 & 1 & 1&1 &1&0&1 \cr
    \textcolor{green}{ 6}& 0 & 0 & 0 & 0 & 1&1 &1&0&0 \cr
    \textcolor{blue}{ 7}& 1 & 1 & 1 & 1 & 1&1 &1&0&1 \cr
    \textcolor{blue}{ 8}& 0 & 0 & 0 & 1 & 0&0 &0&1&0 \cr
    \textcolor{blue}{ 9}& 0 & 0 & 0 & 0 & 1&0 &1&0&1 \cr
  }
  }}}
 \ \ \ \rightarrow \ \ \ \ \ \   Candidates = \{ \textcolor{red}{2}, \ \textcolor{red}{3}, \ \textcolor{green}{4},\ \textcolor{green}{6},\ \textcolor{blue}{8}, \ \textcolor{blue}{9} \}
\]  
\end{minipage}

\begin{minipage}[c]{0\textwidth}
$S.\ \mathbf{2}$
  \[
  \left(\begin{array}{ccccccccc}
    1 & 1 & 1 & 1 & 1 &0 &1&0&0 \\
    \rowcolor{red!20}
    \tikzmark{left}{1} & 1 & 0 & 0 & 1 &0 &1&0&\tikzmark{right}{0}  \\
    \rowcolor{red!20}
    {1} & 0 & 1 & 0 & 1&0 &1&0&{0}   \\
    \rowcolor{green!20}
   1 & 0 &0 & 1 & 1&0 &1&1&0  \\
   1 & 1 & 1 & 1 & 1&1 &1&0&1 \\
    \rowcolor{green!20}
    0 & 0 & 0 & 0 & 1&1 &1&0&0 \\
    1 & 1 & 1 & 1 & 1&1 &1&0&1 \\
    \rowcolor{blue!20}
    0 & 0 & 0 & 1 & 0&0 &0&1&0 \\
    \rowcolor{blue!20}
    0 & 0 & 0 & 0 & 1&0 &1&0&1 
  \end{array}\right)
  \Highlight[first]
  \underbrace{\underset{\{ \textcolor{red}{2}, \  (p, \ q) \   \in \mathcal{P}^{2} \}: \ 
 \mathsmaller{\mathsmaller{\mathsmaller{\bordermatrix{
  	& \textcolor{red}{ 2} & \textcolor{green}{p} &\textcolor{blue}{q} \cr
  	\textcolor{red}{ 2} &1 & &\cr
  	\textcolor{green}{ p} &&1&\cr
  	\textcolor{blue}{ q}&&&1\cr
  	}}}}=\mathbf{I}_{3}}{ \begin{array}{ccccc}
  	\tikzmark{left}{ \{ \textcolor{green}{4}, \ \ \ }  & \  \textcolor{green}{6}, & \textcolor{blue}
  	{8}, \ &  \tikzmark{right}{\ \ \ \textcolor{blue}{9}  \} }\\
  	\end{array} }}_{\mathlarger{\textcolor{red}{M_{2}}} \ }
  	\Highlight[second]
\]

\tikz[overlay,remember picture] {
  \draw[->,thick,red,dashed] (first) -- (second) node [pos=0.6,above] {$\mathcal{P}^{2}$};
  \node[above of=first] {};
  \node[above of=second] {};
}


\end{minipage}
\begin{minipage}[c]{1.8\textwidth}
\vspace{0.5cm}
  \[
  \left(\begin{array}{ccccccccc}
    1 & 1 & 1 & 1 & 1 &0 &1&0&0 \\
    \rowcolor{red!20}
    {1} & 1 & 0 & 0 & 1 &0 &1&0&{0}  \\
    \rowcolor{red!20}
    \tikzmark{left}{1} & 0 & 1 & 0 & 1&0 &1&0&\tikzmark{right}{0}   \\
    \rowcolor{green!20}
   1 & 0 &0 & 1 & 1&0 &1&1&0  \\
   1 & 1 & 1 & 1 & 1&1 &1&0&1 \\
    \rowcolor{green!20}
    0 & 0 & 0 & 0 & 1&1 &1&0&0 \\
    1 & 1 & 1 & 1 & 1&1 &1&0&1 \\
    \rowcolor{blue!20}
    0 & 0 & 0 & 1 & 0&0 &0&1&0 \\
    \rowcolor{blue!20}
    0 & 0 & 0 & 0 & 1&0 &1&0&1 
  \end{array}\right)
  \Highlight[first]
  \underbrace{\underset{\{ \textcolor{red}{3}, \ (p, \ q) \   \in \mathcal{P}^{3} \}:\ 
 \mathsmaller{\mathsmaller{\bordermatrix{
  	& \textcolor{red}{ 3} & \textcolor{green}{p} &\textcolor{blue}{q} \cr
  	\textcolor{red}{ 3} &1 & &\cr
  	\textcolor{green}{ p} &&1&\cr
  	\textcolor{blue}{ q}&&&1\cr
  	}}}=\mathbf{I}_{3}}{ \begin{array}{ccccc}
  	\tikzmark{left}{ \{ \textcolor{green}{4}, \ \ \ }  & \  \textcolor{green}{6}, & \textcolor{blue}
  	{8}, \ &  \tikzmark{right}{\ \ \ \textcolor{blue}{9}  \} }\\
  	\end{array}}}_{\mathlarger{\textcolor{red}{M_{3}}} \  }
  	\Highlight[second]
\]

\tikz[overlay,remember picture] {
  \draw[->,thick,red,dashed] (first) -- (second) node [pos=0.6,above] {$\mathcal{P}^{3}$};
  \node[above of=first] {};
  \node[above of=second] {};
}


\end{minipage}

\begin{minipage}[c]{0\textwidth}
\[
\left(\begin{array}{ccccccccc}
    1 & 1 & 1 & 1 & 1 &0 &1&0&0 \\
    \rowcolor{red!20}
    {1} & 1 & 0 & 0 & 1 &0 &1&0&{0}  \\
    \rowcolor{red!20}
    {1} & 0 & 1 & 0 & 1&0 &1&0&{0}   \\
    \rowcolor{green!20}
   \tikzmark{left}{1} & 0 &0 & 1 & 1&0 &1&1& \tikzmark{right}{0}  \\
   1 & 1 & 1 & 1 & 1&1 &1&0&1 \\
    \rowcolor{green!20}
    0 & 0 & 0 & 0 & 1&1 &1&0&0 \\
    1 & 1 & 1 & 1 & 1&1 &1&0&1 \\
    \rowcolor{blue!20}
    0 & 0 & 0 & 1 & 0&0 &0&1&0 \\
    \rowcolor{blue!20}
    0 & 0 & 0 & 0 & 1&0 &1&0&1 
  \end{array}\right)
  \Highlight[first]
\underbrace{\underset{\{ \textcolor{green}{4},\ (p, \ q) \   \in \mathcal{P}^{4} \}:\ 
 \mathsmaller{\mathsmaller{\bordermatrix{
  	& \textcolor{green}{ 4} & \textcolor{red}{p} &\textcolor{blue}{q} \cr
  	\textcolor{green}{ 4} &1 & &\cr
  	\textcolor{red}{ p} &&1&\cr
  	\textcolor{blue}{ q}&&&1\cr
  	}}}=\mathbf{I}_{3}}{ \begin{array}{cccc}
  	\tikzmark{left}{ \{ \textcolor{red}{2}, \ \ \ } & \ \textcolor{red}{3},  &  \tikzmark{right}{\ \ \ \textcolor{blue}{9}  \} }\\
  	\end{array}}}_{\mathlarger{\textcolor{green}{M_{4}}}  \ }
  	\Highlight[second]
\]

\tikz[overlay,remember picture] {
  \draw[->,thick,red,dashed] (first) -- (second) node [pos=0.6,above] {$\mathcal{P}^{4}$};
  \node[above of=first] {};
  \node[above of=second] {};
}


\end{minipage}
\begin{minipage}[c]{1.8\textwidth}
\[
\left(\begin{array}{ccccccccc}
    1 & 1 & 1 & 1 & 1 &0 &1&0&0 \\
    \rowcolor{red!20}
    {1} & 1 & 0 & 0 & 1 &0 &1&0&{0}  \\
    \rowcolor{red!20}
    {1} & 0 & 1 & 0 & 1&0 &1&0&{0}   \\
    \rowcolor{green!20}
   1 & 0 &0 & 1 & 1&0 &1&1&0  \\
   1 & 1 & 1 & 1 & 1&1 &1&0&1 \\
    \rowcolor{green!20}
    \tikzmark{left}{0} & 0 & 0 & 0 & 1&1 &1&0& \tikzmark{right}{0} \\
    1 & 1 & 1 & 1 & 1&1 &1&0&1 \\
    \rowcolor{blue!20}
    0 & 0 & 0 & 1 & 0&0 &0&1&0 \\
    \rowcolor{blue!20}
    0 & 0 & 0 & 0 & 1&0 &1&0&1 
  \end{array}\right)
  \Highlight[first]
 \underbrace{\underset{\{ \textcolor{green}{6},\ (p, \ q) \   \in \mathcal{P}^{6} \}:\ 
 \mathsmaller{\mathsmaller{\bordermatrix{
  	& \textcolor{green}{ 6} & \textcolor{red}{p} &\textcolor{blue}{q} \cr
  	\textcolor{green}{ 6} &1 & &\cr
  	\textcolor{red}{ p} &&1&\cr
  	\textcolor{blue}{ q}&&&1\cr
  	}}}=\mathbf{I}_{3}}{\begin{array}{ccccc}
  	\tikzmark{left}{ \{ \textcolor{red}{1}, \ \ \ } & \ \textcolor{red}{2}, &\ \textcolor{red}{3},  \ & \  \textcolor{blue}{8}, \  &  \tikzmark{right}{\ \ \ \textcolor{blue}{9}  \} }\\
  	\end{array}}}_{\mathlarger{\textcolor{green}{M_6}}  \   }
  	\Highlight[second]
\]

\tikz[overlay,remember picture] {
  \draw[->,thick,red,dashed] (first) -- (second) node [pos=0.6,above] {$\mathcal{P}^{6}$};
  \node[above of=first] {};
  \node[above of=second] {};
}

\end{minipage}

\begin{minipage}[c]{0\textwidth}
  \[
\left(\begin{array}{ccccccccc}
    1 & 1 & 1 & 1 & 1 &0 &1&0&0 \\
    \rowcolor{red!20}
    {1} & 1 & 0 & 0 & 1 &0 &1&0&{0}  \\
    \rowcolor{red!20}
    {1} & 0 & 1 & 0 & 1&0 &1&0&{0}   \\
    \rowcolor{green!20}
   1 & 0 &0 & 1 & 1&0 &1&1&0  \\
   1 & 1 & 1 & 1 & 1&1 &1&0&1 \\
    \rowcolor{green!20}
    0 & 0 & 0 & 0 & 1&1 &1&0&0 \\
    1 & 1 & 1 & 1 & 1&1 &1&0&1 \\
    \rowcolor{blue!20}
    \tikzmark{left}{0} & 0 & 0 & 1 & 0&0 &0&1&\tikzmark{right}0 \\
    \rowcolor{blue!20}
    0 & 0 & 0 & 0 & 1&0 &1&0&1 
  \end{array}\right)
  \Highlight[first]
  \underbrace{ \underset{\{ \textcolor{blue}{8},\ (p, \ q) \  \in \mathcal{P}^{8} \}:\ 
 \mathsmaller{\mathsmaller{\bordermatrix{
  	& \textcolor{blue}{ 8} & \textcolor{red}{p} &\textcolor{green}{q} \cr
  	\textcolor{blue}{ 8} &1 & &\cr
  	\textcolor{red}{ p} &&1&\cr
  	\textcolor{green}{ q}&&&1\cr
  	}}}=\mathbf{I}_{3}}{ \begin{array}{ccccc}
  	\tikzmark{left}{ \{ \textcolor{red}{1}, \ \ \ } & \ \textcolor{red}{2}, & \ \textcolor{red}{3}, & \  \textcolor{green}{5} \ & \ \    \tikzmark{right}{\ \ \ \textcolor{green}{6} \} }\\
  	\end{array}}}_{ \mathlarger{\textcolor{blue}{M_8} \  }}
  	\Highlight[second]
\]

\tikz[overlay,remember picture] {
  \draw[->,thick,red,dashed] (first) -- (second) node [pos=0.6,above] {$\mathcal{P}^{8}$};
  \node[above of=first] {};
  \node[above of=second] {};
}

\end{minipage}
\begin{minipage}[c]{1.8\textwidth}
\vspace{0.5cm}
  \[
\left(\begin{array}{ccccccccc}
    1 & 1 & 1 & 1 & 1 &0 &1&0&0 \\
    \rowcolor{red!20}
    \tikzmark{left}{1} & 1 & 0 & 0 & 1 &0 &1&0&\tikzmark{right}{0}  \\
    \rowcolor{red!20}
    {1} & 0 & 1 & 0 & 1&0 &1&0&{0}   \\
    \rowcolor{green!20}
   1 & 0 &0 & 1 & 1&0 &1&1&0  \\
   1 & 1 & 1 & 1 & 1&1 &1&0&1 \\
    \rowcolor{green!20}
    0 & 0 & 0 & 0 & 1&1 &1&0&0 \\
    1 & 1 & 1 & 1 & 1&1 &1&0&1 \\
    \rowcolor{blue!20}
    {0} & 0 & 0 & 1 & 0&0 &0&1& {0} \\
    \rowcolor{blue!20}
    \tikzmark{left}{0} & 0 & 0 & 0 & 1&0 &1&0&\tikzmark{right}{1} 
  \end{array}\right)
  \Highlight[first]
 \underbrace{ \underset{\{ \textcolor{blue}{9},\ (p, \ q) \  \in \mathcal{P}^{9} \}:\ 
 \mathsmaller{\mathsmaller{\bordermatrix{
  	& \textcolor{blue}{ 9} & \textcolor{red}{p} &\textcolor{green}{q} \cr
  	\textcolor{blue}{ 9} &1 & &\cr
  	\textcolor{red}{ p} &&1&\cr
  	\textcolor{green}{ q}&&&1\cr
  	}}}=\mathbf{I}_{3}}{\begin{array}{ccccc}
  	\tikzmark{left}{ \{ \textcolor{red}{1}, \ \ \ } & \ \textcolor{red}{2}, &\textcolor{red}{3}, & \textcolor{green}{4},  \  \tikzmark{right}{\ \ \ \textcolor{green}{6}  \} }\\
  	\end{array}} }_{\mathlarger{\textcolor{blue}{M_9}}  \     }
  	\Highlight[second]
\]

\tikz[overlay,remember picture] {
  \draw[->,thick,red,dashed] (first) -- (second) node [pos=0.6,above] {$\mathcal{P}^{9}$};
  \node[above of=first] {};
  \node[above of=second] {};
}

\end{minipage}

\begin{minipage}[c]{1\textwidth}
$S.\ \mathbf{3}$

$$\mathbf{i_{1}}=\textcolor{red}{\textcircled{2}} \ if \ \textcolor{red}{M_2}> \textcolor{red}{M_3}
\mbox{ or }\textcolor{red}{\textcircled{3}} \ if \ \textcolor{red}{M_3}> \textcolor{red}{M_2} $$
$$\mathbf{i_{2}}=\textcolor{green}{\textcircled{4}} \ if \ \textcolor{green}{M_4}> \textcolor{green}{M_6}
\mbox{ or } \textcolor{green}{\textcircled{6}} \ if \ \textcolor{green}{M_6}>\textcolor{green}{M_4}$$
$$\mathbf{i_{3}}=\textcolor{blue}{\textcircled{8}} \ if \ \textcolor{blue}{M_8}> \textcolor{blue}{M_9}   
\mbox{ or }\textcolor{blue}{\textcircled{9}} \ if \ \textcolor{blue}{M_9}> \textcolor{blue}{M_8}$$
\end{minipage}
\caption{Graphical scheme of the MUS algorithm for $K=3$ and precision parameter $\bar{m}=2$.  $S. \ \mathbf{1} $ chooses the candidate maxima, the two units for each group with the greatest number of zeros. $S. \ \mathbf{2} $ identifies for each candidate the subsets $\mathcal{P}$ of the units which belong to a different group (than the candidate) and that have a zero in correspondence of it; then builds all the identity matrices of rank three which contain the candidates. $S. \ \mathbf{3} $ detects the maxima as the three units---one for each group---that appear the greatest number of times in an identity matrix.}
\end{figure}

Before introducing mathematical details, let us denote with $i_{1},...,i_{K}$ a set of $K$ maxima units and with $S_{i_{1},...,i_{K}} $ the submatrix of $C$ 
containing only the rows and columns corresponding to the maxima. The main steps of the algorithm are summarized below.

\begin{description}
\item[(i)] For every group $k, \  k = 1,...,K$ find the \textit{candidate maxima} units $j^{1}_{k},..,j^{\bar{m}}_{k}$ within matrix $C$, i.e.
the units in group $k$ with the greater number of zeros corresponding to the units
of the other $K-1$ groups, where $\bar{m}$ is a \textit{precision parameter} fixed in advance. Let  $\mathcal{P}^{h}_{k}, \ h=1,...,\bar{m}, \ k=1,...,K$ be the entire subset of units belonging to the remaining $K-1$ groups which have a zero in  $j^{h}_{k}$, that is
\begin{equation*}
\mathcal{P}^{h}_{k}= \{ j_{l}, \ l \ne k : C_{({j^{h}_{k}, \  j_{l})}}=0 \}, \ \  \ h=1,...,\bar{m}, \ k=1,...,K
\end{equation*}

where $ C_{({j^{h}_{k}, \  j_{l}})} $ is the  element  $(j^{h}_{k},j_{l})$ of the matrix. We collect a total number of $\bar{m} K$ candidate maxima, $\bar{m}$ for every group.
\item[(ii)] For each of these $\bar{m} K$ units, count the number of distinct identity submatrices of $C$ which contain them, constructed by taking a given candidate $h$ and $K-1$ elements of the corresponding set $\mathcal{P}^{h}_{k}$. Let us denote this quantity with
\begin{equation}
M_{j^{h}_{k}}=  \# \{ S_{j_{1},...,j_{k-1},\ j^{h}_{k},\ j_{k+1},...,j_{K}}| \ j_{i} \in \mathcal{P}^{h}_{k}, \ i=1,...,k-1,k+1,...,K \}.
\label{eq:M}
\end{equation}

\item[(iii)] For each group $k,\, k = 1,...,K$, select the unit which yields the greatest number of identity matrices of rank $K$. In mathematical terms

\begin{equation}
i_{k} =\underset{j^{h}_{k} \in \{j^{1}_{k},...,j^{\bar{m}}_{k} \}}{\mbox{argmax}} \  M_{j^{h}_{k}},\,  \ \ \ h=1,...\bar{m}, \ \  k=1,...,K.
\label{eq:maxima}
\end{equation}
\end{description}

The choice of $\bar{m}$ is relevant in terms of the algorithm performance. This parameter is a sort of benchmark for the size of the $K$ subsets where the algorithm searches for the $K$ maxima units: the smaller is this value, the less comprehensive is the search. Conversely, a bigger value enhances the possibility to build a larger set $\mathcal{P}^{h}_{k}$ and obtain a more accurate result. In Sect.~\ref{sec:sim_study} we deal with this issue and we consider different choices for the precision parameter $\bar{m}$.

\section{Simulation study}
\label{sec:sim_study}

The task of this section is to investigate the performance of the MUS algorithm and its sensitivity to the choice of $N$ and $\bar{m}$, for a fixed $K$, which is determined by some clustering technique or a given grouping of the units. To this aim, we simulate a symmetric $N \times N$ matrix $C$ where the element $(i,j)$ is drawn from a Bernoulli distribution with parameter $p$. As mentioned in Sect.~\ref{sec:meth}, the $i$-th row's index, $i, i=1,...,N$, is associated to a statistical unit of interest and each unit is here randomly assigned to group $k$, $k=1,...K$, with probability $1/K$.
We consider three different values of $p$, i.e. $p=0.8, 0.5, 0.2$. 

 Tables~\ref{tab:p},~\ref{tab:p2} and \ref{tab:p3} display the maxima units and the corresponding CPU times in seconds (in brackets) according to the considered scenarios. As expected, the procedure is sensitive to the choices of input weights,  both in terms of units selection and computational times. 

The first issue one may immediately notice is that, regardless of the weights used for generating data, the computational burden rises dramatically when $K>3$. Especially when $N=1000$, the CPU time is huge if compared to the time spent in the same framework---same $\bar{m}$ and weights--- for $K=2$ or $K=3$. As the probability $p$ decreases (from $0.8$ to $0.2$) the number of zeros becomes larger and, consequently, the CPU time required keep growing regardless of the values of $N$ and $\bar{m}$.
A second remark is that, by fixing $N$ and $K$, the choice of the precision parameter $\bar{m}$ does not seem to affect significantly the performance of the procedure: as $\bar{m}$ increases, there is limited variation in the units detection and the difference in the required time between $\bar{m}=1$ and $\bar{m}=20$ remains relatively small, as can be seen from Tables~\ref{tab:p}, \ref{tab:p2} and \ref{tab:p3}. This is suggesting that even the choice of a small precision parameter---e.g. $\bar{m}=5 $---may be accurate enough for detecting the maxima.

\begin{table}[H]
\centering
\begin{tabular}{c|c|c|c|c|c|}
&& $K=2$ & $K=3$ & $K=4$ \\
\hline
 \multirow{4}{*}{$N=100$} & $\bar{m}=1$ & $43, \  14  \ \mathbf{(<0.1)}$& $ - \ - \  - \ \mathbf{(<0.1)}$ &$18,\ 34,\ 41,\ 88 \ \mathbf{( <0.1 )}$ \\
& $\bar{m}=5$ & $16 , \  14 \mathbf{(<0.1)}$ & $10, \ 49, \ 96 \ \mathbf{(<0.1)}$ & $37,\ 78,\ 17,\ 69 \ \mathbf{( 0.16 )}$ \\
 &$\bar{m}=10$ & $16, \  14 \ \mathbf{(<0.1)}$ & $10, \ 49, \ 96 \ \mathbf{(<0.1)}$ &  $37,\ 78,\ 65,\ 69 \ \mathbf{( 0.25 )}$\\
 & $\bar{m}=20$ & $ 16, 14 \ \mathbf{(<0.1)}$ & $10, \ 49, \ 96 \ \mathbf{(<0.1)}$& $37,\ 78,\ 65,\ 69 \ \mathbf{( 0.43 )}$\\ \hline
\multirow{4}{*}{$N=500$} & $\bar{m}=1$ & $ - - \mathbf{(0.56)} $& $ - \ - \  - \ \mathbf{(0.63)}$ & $27,\ 44,\ 59,\ 263, \ \mathbf{( 2.3)}$\\
 &$\bar{m}=5$ & $ 183,\  125 \  \mathbf{(0.66)}$ & $346, \  373, \   500 \ \mathbf{(0.68)}$ & $394,\ 44,\ 59,\ 263, \ \mathbf{( 9.7)}$\\
 & $\bar{m}=10$ & $183, \  125 \ \mathbf{(0.64)}$ & $ 399, \  373, \   500 \ \mathbf{(0.94)}$ & $394,\ 44,\ 59,\ 263, \ \mathbf{( 17.6)}$\\
 & $\bar{m}=20$ & $ 183, \  125 \  \mathbf{(0.72)}$ &  $ 399, \  373, \   500 \ \mathbf{(1.22)}$ &$394,\ 44,\ 59,\ 263, \ \mathbf{( 32.71)}$\\ \hline
\multirow{4}{*}{$N=1000$} & $\bar{m}=1$ & $ -\   - \  \mathbf{(2.32)}$& $ - \ - \  - \  \mathbf{(2.40)}$ & $350,\ 825,\ 916,\ 204 \ \mathbf{( 10.9)}$\\
 &$\bar{m}=5$ &  $654, \   94 \  \mathbf{(2.49)}$ & $909, \   499,\   868 \ \mathbf{(3.02)} $ &$381,\ 849,\ 684,\ 204 \ \mathbf{( 44.0)}$\\
 & $\bar{m}=10$ &$ 654,\ 94 \ \mathbf{(2.62)}$& $909, \   499,\   868  \ \mathbf{(3.52)} $ & $381,\ 849,\ 684,\ 488 \ \mathbf{(81.7)}$\\
 & $\bar{m}=20$ & $654, \   96 \  \mathbf{(2.99)}$& $909, \   382,\   868 \ \mathbf{(4.62)}$ & $381,\ 849,\ 748,\ 488 \ \mathbf{(152.82)}$\\ \hline
\end{tabular}
\caption{MUS algorithm's maxima and computational times (in brackets) according to $K=2,3,4,\ \ N=100,500,1000, \  \bar{m}=1,5,10,20 $. Bernoulli data $0,1$ generated with weights $p=0.8$.}
\label{tab:p}
\end{table}

\begin{table}[H]
\centering
\begin{tabular}{c|c|c|c|c|c}
&& $K=2$ & $K=3$ & $K=4$  \\
\hline
 \multirow{4}{*}{$N=100$} & $\bar{m}=1$ & $48, \ 86 \  \mathbf{(<0.1)}$ & $ - \ - \  - \ \mathbf{(<0.1)} $& $32,\ 62,\  38,\ 55 \ \mathbf{( 0.44 )}$ \\
& $\bar{m}=5$ & $48, \ 61 \  \mathbf{(<0.1)}$ & $15,\    33,\     5, \ \mathbf{(0.12)}$ & $50,\ 62,\  89,\ 55 \ \mathbf{( 1.99 )}$ \\
 &$\bar{m}=10$ & $ 48, \ 61 \ \mathbf{(<0.1)}$ & $15,\    62,\     5 \ \mathbf{(0.14)}$ & $50,\ 62,\  90,\ 55 \ \mathbf{( 3.36 )}$\\
 & $\bar{m}=20$ & $48, \ 61 \ \mathbf{(0.10)}$ & $15,\    62,\     5 \ \mathbf{(0.13)} $ & $50,\ 62,\  90,\ 55 \ \mathbf{( 1328.73 )}$\\ \hline
\multirow{4}{*}{$N=500$} & $\bar{m}=1$& $-$ $- \ \mathbf{(1.61)}$& $ - \ - - \  (\mathbf{1.67)}$& $10,\ 11,\ 90,\ 488 \ \mathbf{( 56.1)}$\\
 &$\bar{m}=5$ & $294, \  242 \ \mathbf{(1.40)}$ & $203,\    213,\     272 \ \mathbf{(2.31)}$ & $273,\ 242,\ 292,\ 383 \ \mathbf{( 159.8)}$\\
 & $\bar{m}=10$ & $ 294, \  242 \  \mathbf{(1.64)} $ & $203,\    213,\     272  \ \mathbf{(3.31)}$ & $273,\ 232,\ 482,\ 383 \ \mathbf{( 311.28)}$ \\
 & $\bar{m}=20$ & $66, \  242 \ \mathbf{(1.78)}$ & $203,\    213,\     272 \ \mathbf{(5.49)} $ & $273,\ 232,\ 29,\ 383 \ \mathbf{( 582.38)}$\\ \hline
\multirow{4}{*}{$N=1000$} & $\bar{m}=1$ & $-$ $-$ (6.64)& $ - \ - \  - \ \mathbf{(7.16)} $& $ 123,\ 964,\ 813,\ 238 \  \mathbf{( 246.28)}$\\
 &$\bar{m}=5$ & $94, \ 405 \  \mathbf{(6.81)}$ & $67,\    995,\     688, \ \mathbf{(9.78)}$ & $ 267,\ 964,\ 813,\ 241 \  \mathbf{( 1208.27)}$\\
 & $\bar{m}=10$ & $94, \  405 \  \mathbf{(7.24)}$& $67,\    995,\     688,\ \mathbf{(12.61)}$ & $ 267,\ 964,\ 813,\ 241 \  \mathbf{( 2326.64)}$\\
 & $\bar{m}=20$ & $398, \  405 \ \mathbf{(8.23)}$ & $67,\    995,\     688, \ \mathbf{(9.58)}$ & $ 267,\ 964,\ 813,\ 241 \  \mathbf{( 4548.47)}$\\ \hline
\end{tabular}
\caption{MUS algorithm's maxima and computational times  (in brackets) according to $K=2,3,4,\ \ N=100,500,1000, \  \bar{m}=1,5,10,20 $. Bernoulli data $0,1$ generated with weights $p=0.5$.}
\label{tab:p2}
\end{table}

\begin{table}[H]
\label{tab:p3}
\centering
\begin{tabular}{c|c|c|c|c|c}
&& $K=2$ & $K=3$ & $K=4$ \\
\hline
 \multirow{4}{*}{$N=100$} & $\bar{m}=1$ & $42, \ 32$ $ \ \mathbf{(<0.1)}$ & $58,\   40,\   63 \ \mathbf{(0.10)}$& $ 24,\ 68,\ 34,\ 89 \ \mathbf{( 3.75 )}$  \\
& $\bar{m}=5$ & $42, \ 32$ $ \ \mathbf{(<0.1)}$ & $81,\   54,\   63 \ \mathbf{(0.21)}$ & $ 24,\ 68,\ 48,\ 58 \ \mathbf{( 8.85 )}$  \\
 &$\bar{m}=10$ & $42, \ 86$ $ \ \mathbf{(0.14)}$ & $87,\   54,\   63 \ \mathbf{(0.25)}$ & $ 24,\ 68,\ 48,\ 58 \ \mathbf{( 16.8 )}$ \\
 & $\bar{m}=20$ & $42, \ 86$ $ \ \mathbf{(0.11)}$ & $87,\   54,\   63 \ \mathbf{(0.43)}$ & $ 24,\ 68,\ 48,\ 58 \ \mathbf{( 1985.01 )}$ \\ \hline
\multirow{4}{*}{$N=500$} & $\bar{m}=1$ & $-$ $-$ $\ \mathbf{(2.40)}$ &$ - \ - \  - \ \mathbf{(2.74)}$ & $371,\ 28,\ 122,\ 60 \ \mathbf{( 189.94)}$\\
 &$\bar{m}=5$ & $326,\ 288\ \mathbf{(2.39)}$ & $290, \ 393, \ 316 \ \mathbf{(4.51)}$& $370,\ 38,\ 202,\ 404 \ \mathbf{( 949.4)}$ \\
 & $\bar{m}=10$ & $326,\ 288 \ \mathbf{(2.65)}$ & $290, \ 393, \ 316 \ \mathbf{(7.20)}$& $370,\ 413,\ 202,\ 196 \ \mathbf{( 1882.93)}$  \\
 & $\bar{m}=20$ & $284,\ 288 \ \mathbf{(3.06)}$ & $375, \ 395, \ 316 \ \mathbf{(10.66)}$& $370,\ 38,\ 202,\ 404 \ \mathbf{( 3685.61)}$ \\ \hline
\multirow{4}{*}{$N=1000$} & $\bar{m}=1$ & $555, \ 892, \ \mathbf{(11.30)}$ & $ - \ - \  - \ \mathbf{(12.25)}$ & $427,\ 452,\ 218,\ 631 \ \mathbf{( 1608.11 )}$\\
 &$\bar{m}=5$ & $434,\  892 \ \mathbf{(11.28)}$ & $ 222, \ 921, \  275 \ \mathbf{(19.42)}$ & $427,\ 493,\ 218,\ 839 \ \mathbf{( 8098.54 )}$\\
 & $\bar{m}=10$ & $434,\  892 \ \mathbf{(11.27)} $ & $ 387, \ 921, \  255 \ \mathbf{(26.72)}$ & $427,\ 493,\ 218,\ 839 \ \mathbf{( 16629.86 )}$\\
 & $\bar{m}=20$ & $434,\  892 \ \mathbf{(12.47)} $ &  $ 387, \ 921, \  255 \ \mathbf{(45.08)}$ & $427,\ 493,\ 218,\ 839 \ \mathbf{( 32230.8 )}$ \\ \hline
\end{tabular}
\caption{MUS algorithm's maxima and computational times  (in brackets) according to $K=2,3,4,\ \ N=100,500,1000, \  \bar{m}=1,5,10,20 $. Bernoulli data $0,1$ generated with weights $p=0.2$.}
\end{table}

\section{Applications}
\label{sec:appl}

\subsection{Identification of pivotal units}
\label{subsec:pivotal}

As broadly explained in \cite{pauli2015relabelling}, the identification of some pivotal units in a Bayesian mixture model with a fixed number of groups may be helpful when dealing with  the label switching problem (\cite{stephens2000dealing}, \cite{jasra2005markov}).


Let $N$ be the number of observations generated from the mixture model. Consider, for instance, the probability of two units being in the same group. Such quantity may be estimated from the MCMC sample and denoted as $\hat{c}_{ij}$. For details, see \cite{pauli2015relabelling}. The $N \times N$ matrix $C$ with elements $\hat{c}_{ij}$ can be seen as a similarity matrix between units. Now, such matrix can be considered as input for some suitable clustering techniques, in order to obtain a partition of the $N$ observations into $K$ groups.

%
%
From such partition, we may be interested in identifying exactly $K$ pivotal units---\textit{pivots}---which are (pairwise)
separated with (posterior) probability one (that is, the posterior probability of any two of
them being in the same group is zero). In fact, as discussed in  \cite{pauli2015relabelling},  the identification of such units allows us to provide a valid solution to the occurrence of label switches.

 In terms of the matrix $C$, the submatrix $S$ with
only the rows and columns corresponding to the maxima $i_{1},...,i_{K}$ found through the MUS, if they exist, following the steps in Sect.~\ref{sec:meth}, is the identity matrix.  It is still worth noticing that the availability of $K$ perfectly separated units is crucial to the procedure, and it can
not always be guaranteed. 

Practically, our interest is in finding units which should be `as far as possible' one from each other according to a well defined distance measure. The more separated they are, the better they represent the  group they belong to.

\begin{figure}
\centering
\includegraphics[scale=0.32]{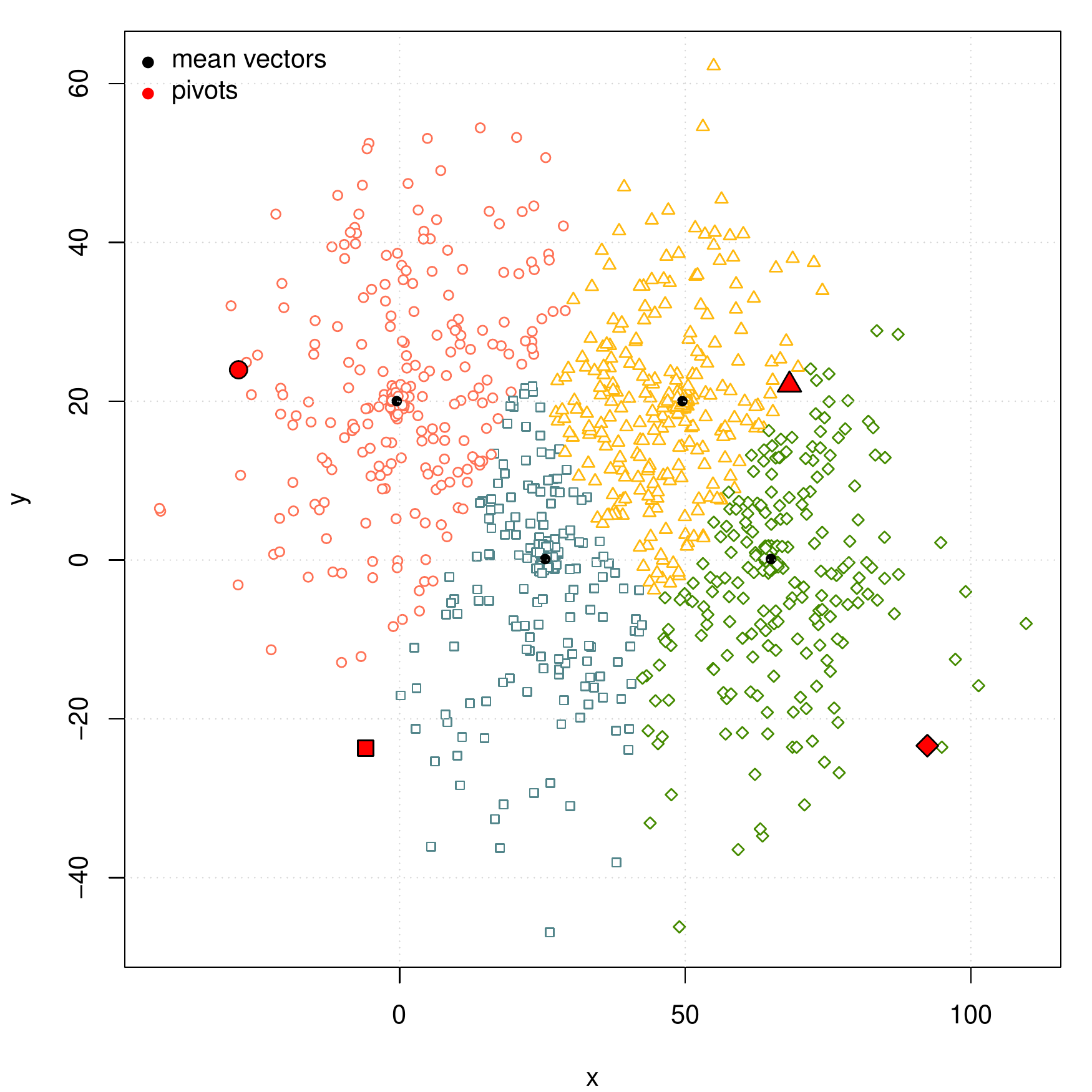}~
\includegraphics[scale=0.32]{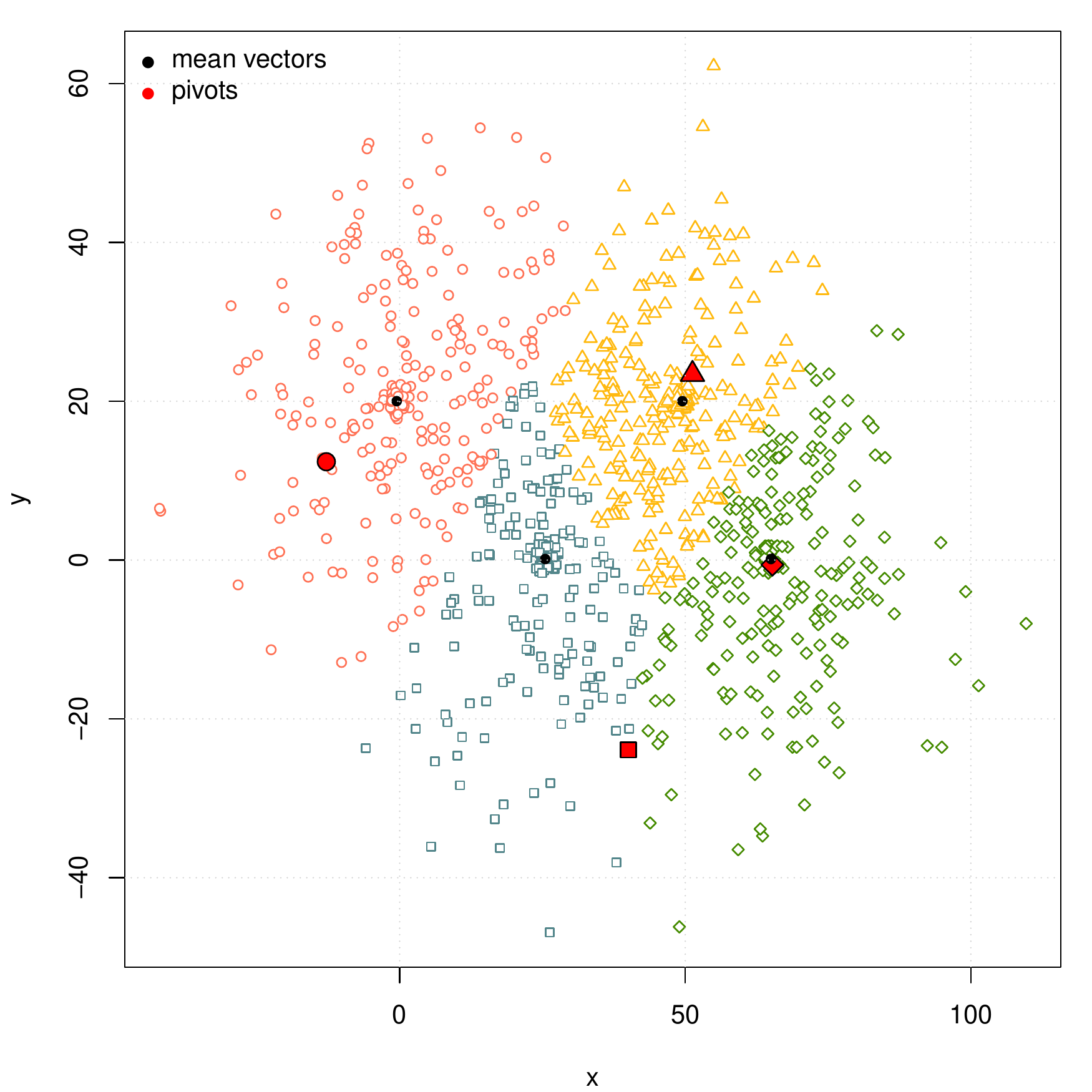}
\caption{Simulated bivariate sample of size $N = 1000$ from a nested Gaussian mixture of mixtures with $K=4$ and input means (in black) $\bm{\mu}_{1}= (25,0),  \bm{\mu}_{2}=(60,0), \  \bm{\mu}_{3}=(0,20), \ \bm{\mu}_{4}=(50,20)$. Groups have been detected through an agglomerative clustering technique. Pivots---i.e. maxima--- found by MUS algorithm (Left) with $\bar{m}=5$ are shown in red and seem well separated in the bi-dimensional space. Pivots found by method $\min_{\bar{i}}(\min_{j\not \in \mathcal{K}_{k}} C_{ (\bar{i} j}))$ are less distant each other (Right). }
\label{fig:pivot}
\end{figure}

Fig.\ref{fig:pivot} shows the pivotal identification of $K=4$ units for a sample of $N=1000$ bivariate data generated according to a nested Gaussian mixture of mixtures with $K$ groups and fixed means. Pivots (red) have been detected through the MUS algorithm and through another alternative method, which aims at searching the most distant units among the members that are farthest apart. We may graphically notice that separation is made more efficient by the MUS algorithm, for which the red points appear quite distant from each other. Moreover, in this specific example the pivotal search is made difficult due to the overlapping  of the $K$ groups.

\subsection{Tennis singular features}
\label{subsec:tennis}

As a simple example we apply our algorithm to a case study 
regarding tennis players.
We collect $N=8$ game's features (hereafter $GF$) for $T=25$ players from  the Wimbledon Tournament 2016\footnote{\url{http://www.wimbledon.com/en_GB/scores/extrastats/index.html}.}, and we assign the following values

$$\begin{cases}
GF_{i,t}=1, & \mbox{ if player } t \mbox{ has  GF } i, \\
GF_{i,t}=0, & \mbox{ if player } t \mbox{ has \textit{not} the GF } i.\\
\end{cases}  $$

Game's features belong to $K=2$ groups, which somehow refer to the attack and  the defence skills  for each player. We denote with the label 1 the first group and with 2 the second group: `First Serve Receiving Points' (2), `Second Serve Receiving Points' (2), `Break Points Won' (1), `Serve Speed' (1), `Aces'(1), `First Serve Points' (1), `Second Serve Points' (1), `Break Points Conversion' (2).

We decide to assign a specific game feature to a given player if this player belongs to the first five positions of that game's feature rank reported by the Wimbledon's website. Hence, let us enumerate the above-mentioned features from one to eight. Our $0-1$ dataset has as many records as players. The problem setting is summarized below.

\begin{table}[H]
\centering
\begin{tabular}{ccccccccc}
&\multicolumn{8}{c}{GF}\\
Player & 1 & 2 &3 & 4 & 5 & 6 & 7 & 8\\
\hline 
Federer &1 &1&0&0&0&0&0&0 \\
Murray &1 & 1 &1&0&0&0&0&0\\
...& ...&...&...&...&...&...&...&\\ 
\end{tabular}
\end{table}

Note that Federer is assigned game's feature one (`First Serve Receiving Points') and two (`Second Serve Receiving Points') only, Murray is assigned game's feature one, two and three (`Break Points Won') and so on.
We define the $N \times N$ symmetric matrix, $C$, in which the generic element $C_{(i,j)}$ is the number of players that have both features $i$ and $j$. 

$$ C= \left(\begin{matrix}
    1  &  5 &   3 &   1  &  3  &  0 &   0  &  0 \\ 
    5  &  1 &   2 &   1  &  2  &  0  &  0   & 0\\ 
    3  &  2  &  1  &  0  &  1   & 0  &  0  &  0 \\ 
     1  &  1 &   0  &  1  &  2  &  0  &  0  &  0 \\ 
    3  &  2 &   1  &  2  &  1  &  0  &  0  &  0\\ 
    0   & 0  &  0  &  0  &  0  &  1  &  0   & 0 \\ 
    0  &  0  &  0 &   0  &  0  &  0  &  1  &  0 \\ 
     0  &  0 &   0 &   0 &   0  &  0 &   0  &  1 \\ 
 \end{matrix} \right)$$
 
 We would like to extract the two `most distant' game's features for the two groups, i.e. the two features in correspondence of which the matrix $C$ is zero more often, and for which $S_{i_{1},i_{2}}$ is an identity matrix. We can notice that rows 6 and 7 of $C$ are full of zeros: this means, for instance, that according to our short dataset $C_{(6,1)}=0$, i.e. the `First Serve Points' ($k=1$) and `First Serve Receiving Points' ($k=2$) do not coexist to any player. Are they the most distant features between the two groups? To answer this question, we run the MUS algorithm by fixing $\bar{m}=3$ and we find the  candidate maxima $ j^{1}_{1}=6, \ j^{2}_{1}=7, \ j^{1}_{2}=8, \ j^{2}_{2}=1 $ and maxima $i_{1}=6, \ \ i_{2}=8 $. Hence we conclude that `First Serve Points' (six) and `Break Points Conversion' (eight) are quite unlikely to belong to the same player.

\section{Conclusions}
\label{sec:concl}

A procedure for detecting small identity submatrices from a $N \times N$ matrix has been proposed. It has been initially considered for application to the pivotal approach in label switching problem in the analysis of Bayesian mixture models. The proposed method is discussed in detail and employed for different practical problems.

 Its efficiency and its sensitivity to parameter choices is investigated through a simulation study, which shows that for a small number of groups the procedure is quite fast.
Moreover, even for small values of the precision parameter $\bar{m}$ the procedure appears quite stable in terms of units indexes, suggesting that a higher value of $\bar{m}$ is often not required. This is also confirmed by the results in Sect.~\ref{sec:appl}.

Further issues for future research are related to the optimization of the proposed algorithm and the definition of suitable indicators for detecting both diagnostic problems inherent to the procedure and goodness of units choice.

%
%

\begin{thebibliography}{99.}%
%



\bibitem{pauli2015relabelling} Egidi, L., Pappad\`{a}, R., Pauli, F., Torelli, N.: Relabelling in Bayesian mixture models by pivotal units. arXiv preprint:1501.05478v2 (2015)
%

\bibitem{fred2002data} Fred, Ana L.N., Jain, Anil K.: Data clustering using evidence accumulation. Pattern Recognition, 2002. Proceedings. 16th International Conference on, \textbf{4}, 276--280 (2002)


\bibitem{jasra2005markov} Jasra, A., Holmes, C.C., Stephens, D.A.: Markov chain Monte Carlo methods and the label switching problem in Bayesian mixture modeling. Statistical Science, 50--67 (2005)


\bibitem{kaiser1974index} Kaiser, Henry F.: An index of factorial simplicity. Psychometrika, \textbf{39(1)}, 31--36 (1974)


\bibitem{kuhn1955hungarian} Kuhn, H.W.: The Hungarian method for the assignment problem. Naval research logistics quarterly, \textbf{2(1-2)}, 83--97 (1955)


\bibitem{munkres1957algorithms} Munkres, J.: Algorithms for the assignment and transportation problems. Journal of the society for industrial and applied mathematics, \textbf{5(1)}, 32--38 (1957)

\bibitem{puolamaki2009bayesian} Puolam{\"a}ki, K., Kaski, S.: Bayesian solutions to the label switching problem. In: Advances in Intelligent Data Analysis VIII, pp. 381--392. Springer (2009)
%
\bibitem{stephens2000dealing} Stephens, M.: Dealing with label switching in mixture models. Journal of the Royal Statistical Society: Series B (Statistical Methodology), \textbf{62(4)}, 795--809 (2000)







%


\end{thebibliography}
%
%

\end{document}